\documentclass[journal]{elsarticle}

\usepackage{hyperref}

\journal{Journal of Computer Methods and Programs in Biomedicine}

\usepackage{tabulary,xcolor}
\usepackage{amsfonts,amsmath,amssymb}
\usepackage{xcolor}
\usepackage{float}

\begin{document}

\begin{frontmatter}

\title{COVIDLite: A depth-wise separable deep neural network with white balance and CLAHE for detection of COVID-19}

\author{Manu Siddhartha\fnref{myfootnote1}}
\address{Department of Computer Science\unskip, International Institute of Information Technology\unskip, Bangalore\unskip, Karnataka, India}
\fntext[myfootnote1]{manusiddhartha.dml11@iiitb.net}

\author{Avik Santra\fnref{myfootnote2}}
\address{Infineon Technologies AG, Neubiberg, Germany}
\fntext[myfootnote2]{avik.santra@gmail.com}

%
%

\begin{abstract}
	
\textbf{Background and Objective:} Currently, the whole world is facing a pandemic disease, novel Coronavirus also known as COVID-19, which spread in more than 200 countries with around 3.3 million active cases and 4.4 lakh deaths approximately. Due to rapid increase in number of cases and limited supply of testing kits, availability of alternative diagnostic method is necessary for containing the spread of COVID-19 cases at an early stage and reducing the death count. For making available an alternative diagnostic method, we proposed a deep neural network based diagnostic method which can be easily integrated with mobile devices for detection of COVID-19 and viral pneumonia using Chest X-rays (CXR) images. 
 
\textbf{Methods:} In this study, we have proposed a method named COVIDLite, which is a combination of white balance followed by Contrast Limited Adaptive Histogram Equalization (CLAHE) and depth-wise separable convolutional neural network (DSCNN). In this method, white balance followed by CLAHE is used as an image preprocessing step for enhancing the visibility of CXR images and DSCNN trained using sparse cross entropy is used for image classification with lesser parameters and significantly lighter in size, i.e., 8.4 MB without quantization. 
 
\textbf{Results:} The proposed COVIDLite method resulted in improved performance in comparison to vanilla DSCNN with no pre-processing. The proposed method achieved higher accuracy of 99.58\% for binary classification, whereas 96.43\% for multi-class classification and out-performed various state-of-the-art methods.

\textbf{Conclusion:} Our proposed method, COVIDLite achieved exceptional results on various performance metrics. With detailed model interpretations, COVIDLite can assist radiologists in detecting COVID-19 patients from CXR images and can reduce the diagnosis time significantly.
  
\end{abstract}


\end{frontmatter}

\section{Introduction}
COVID-19 started to spread since December 2019, when a large number of pneumonia cases of unknown cause occurred in Wuhan, Hubei, China, whose clinical characteristics much resembled with viral pneumonia. Detailed analysis by observing samples indicated a novel virus named as coronavirus or COVID-19. Since then, the number of cases increased extremely fast, with death count increased at an alarming rate. The critical observation in casualty cases from COVID-19 was that all the patients had developed into severe pneumonia. The primary strategy for containing its spread and reducing the death count is early diagnosis of COVID-19 cases so that patients can be timely quarantined, which prevents them from spreading the virus to others ~\cite{adhikari2020epidemiology}. The first and the most common diagnostic test for diagnosing COVID-19 is a real-time reverse transcription-polymerase chain reaction (RT-PCR). However, the RT-PCR test has three significant issues. First, it is a time-consuming process ~\cite{yi2020covid}; second, it has lower sensitivity of 60\%{\textendash}70\%, ~\cite{udugama2020diagnosing}, i.e., around 30\% of the cases where the test diagnosed negative to actual positive cases is a critical issue; third, it requires the availability of commercial kits ~\cite{udugama2020diagnosing}. However, in those misdiagnosed cases, symptoms can be detected when examined through radiological images ~\cite{kanne2020essentials} ~\cite{xie2020chest}. So, other alternative diagnostic tests that are relatively accurate with higher sensitivity and faster in terms of test results are significant in the early diagnosis of COVID-19 patients. According to some of the researchers, combined features of radiological images with other laboratory tests may help in early-stage identification of COVID-19 ~\cite{kong2020chest} ~\cite{lee2020covid} ~\cite{shi2020radiological} ~\cite{zhao2020relation} ~\cite{li2020coronavirus}. In some of the studies, researchers found a useful pattern in chest CT and CXR images, which were very crucial in the identification of COVID-19   ~\cite{kong2020chest} ~\cite{zhao2020relation} ~\cite{yoon2020chest}. Chest CT is more sensitive to COVID-19, but we have employed CXR in this research due to its easier availability, lower cost, and lesser waiting time for results.

Convolutional Neural Networks (CNNs) have recently shown to exceed the medical practitioner's performance in diagnosing pneumonia from CXR images due to advancements in processing power and the availability of large image datasets. In ~\cite{stephen2019efficient}, researchers have used seven layers of deep convolutional neural networks (DCNNs) for detecting pneumonia from CXR images. In this method, they have employed the ReLU activation function between hidden layers and a sigmoid activation function for performing the classification task. They have achieved higher validation accuracy for the image size of 200 x 200. In ~\cite{chhikara2020deep}, a DCNN with transfer learning used to detect pneumonia from CXR images. In this method, researchers have used advanced image processing techniques such as gamma correction, equalization, filtering, and compression to enhance image quality, which improved the overall performance of their model.

Recently, a significant amount of research was conducted for diagnosing COVID-19 from CXR images using DCNNs, and have shown exceptional results. ~\cite{sethy2020detection} have used a transfer learning approach for detecting COVID-19 from CXR images. They have employed Support Vector Machine for extracting deep features, and pre-trained network ResNet-50 for classification, resulting in higher mean accuracy of 95.33\%. In another study ~\cite{loey2020within}, researchers have used generative adversarial network (GAN) to increase training images of four class types COVID-19, normal, viral pneumonia, and bacterial pneumonia, whereas pre-trained networks such as Alexnet, Googlenet, and Restnet18 used for the classification task. They have increased the size of their existing dataset to 30 times, which resulted in higher accuracy and sensitivity of the models. ~\cite{hemdan2020covidx}  proposed a new framework named COVID X-Net capable of detecting COVID-19 from CXR images using seven pre-trained networks namely, VGG19, DenseNet121, ResNetV2, InceptionV3, InceptionResNetV2, Xception and MobileNetV2. 

In this paper, we have developed a hybrid method comprising a deep CNN model based on depth-wise separable convolutions and white balance followed by CLAHE for image enhancement as a data preprocessing step. In previous studies, different novel data preprocessing techniques proved useful in many applications such as hand gesture recognition using radar systems ~\cite{hazra2018robust}, hand gesture recognition using ultrasound imaging and speech recognition ~\cite{picone1993signal} among others. DSCNNs first used in Xception Net has shown exceptional results in the classification task of natural images. However, in the case of medical images, much adjustment is required as medical images can be available in 3 dimensions RGB, 4 dimensions, or 2-dimensions with grayscale, whereas natural images are mostly available in 2-dimensional RGB formats. The other major difference is in terms of intensity variations, furthermore natural images are generally recognized by edges, basic shape, correlation among neighboring pixels etc., whereas each pixel intensity or brightness level are not relevant feature for their recognition. However, in the case of medical images, specifically CXR images, each pixel intensity is essential in locating affected regions within the image, which is significantly crucial for detecting abnormalities within CXR images. Thus, proper image preprocessing is essential for building a robust and accurate model for detecting abnormalities in medical images.

\subsection{Contributions and Outline}

The contribution of this study can be summarized as follows:
  
\begin{enumerate}
  \item \relax A novel approach comprising of white balance with CLAHE followed by novel DSCNN model is proposed for the detection of COVID-19 from CXR images,
  \item \relax The proposed method is shown to outperform traditional DSCNN without preprocessing and other recent state-of-the-art methods in terms of accuracy in detecting COVID-19 in both 2-class and 3-class classification tasks,
  \item \relax The proposed method is significantly lighter in terms of model size making it favorable for deploying as a web app for generating real-time diagnosis of COVID-19 in inaccessible areas,
  \item \relax We analyze the intepretability of the proposed model using Grad-CAM, LIME, and saliency maps, which in turn can provide key insights to radiologists for accurate diagnosis of COVID-19.
\end{enumerate}

The paper is organized as follows. In section 2, we detailed about depth-wise separable CNNs and the method employed in this research. Further, we also presented the architecture of the proposed method. In section 3, we describe the dataset utilized in this research, along with performance measures used for evaluating our method. In section 4, we have presented our results with a visual interpretation of our model's prediction and discussed the previous method used for the detection of COVID-19 and their comparison with the proposed method. Section 5 provides the conclusion of this research.

\section{Methodology}

\subsection{Depth-wise separable CNN (DSCNN)}

\begin{figure}[!htb]
	\centering
	{\includegraphics[width=\linewidth]{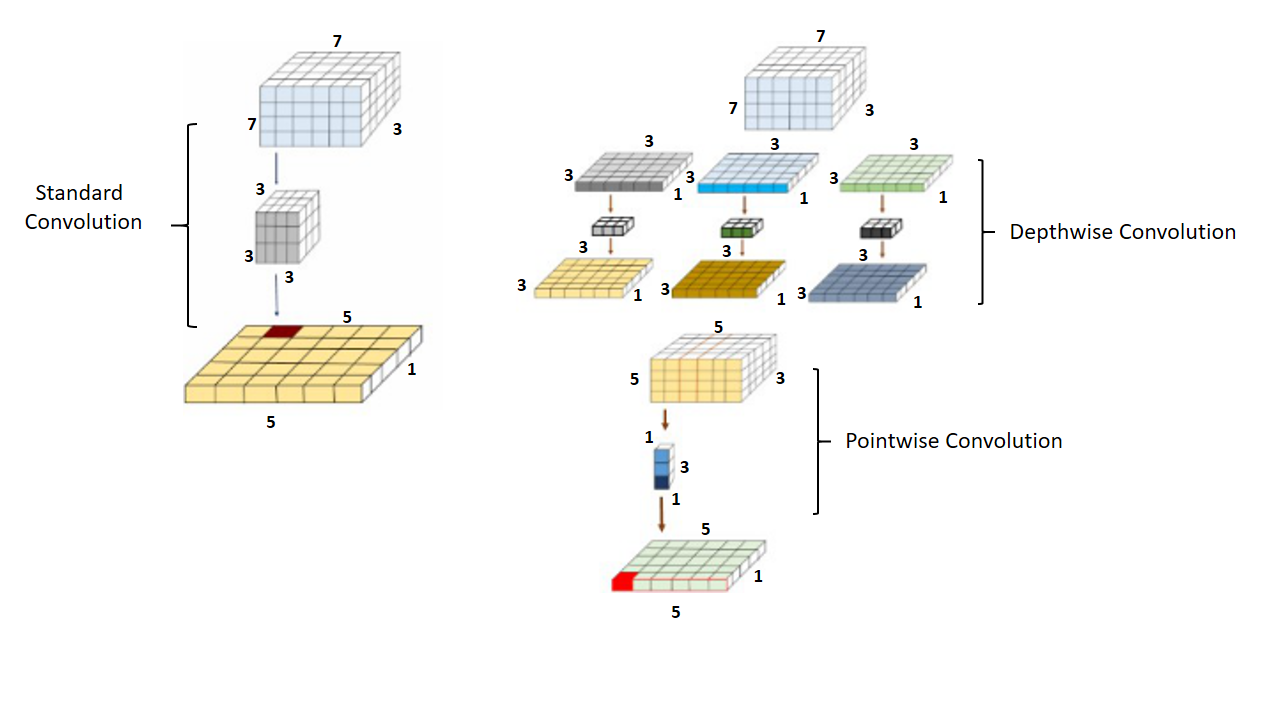} }
	\caption{Comparison of (left) standard convolution and (right) depth seperable convolution operation. }
 	\label{f-3b1de2ef29c0}
\end{figure}

Convolutional Neural Networks (CNNs) inspired from the working of human brain's visual cortex. One of the key reason for the success of CNN is the availability of large training dataset, also their ability to learn distinctive features implicitly from the input image. However, deep CNNs come with the drawback of large memory size and compute requirements, which can be deterrent for real-time applications such as on embedded platform or web application. Further, DCNNs with large parameters require large dataset to avoid overfitting, which is a problem for medical imaging, where datasets are limited. One of the solution for the above problem is depth-wise separable convolution (DSC) first proposed by Chollet ~\cite{chollet2017xception}, which divides the regular convolution operation into two separate operations depth-wise or spatial convolution and sequential point-wise convolution, as shown in Fig. \ref{f-3b1de2ef29c0}. The depth-wise convolution operation applies a 2D filter to each input channel. Since a single filter is applied to each input channel, it reduces the number of computations significantly as only the input channel corresponding to each filter is computed. Further, $1 \times 1$ convolution or point-wise convolution is used to aggregate the results of depth-wise convolution for creating a new feature map, as shown in Fig. \ref{f-3b1de2ef29c0}(b). Thus resulting in significant reduction in parameters and computation cost, making them favorable candidate for real-time inference on mobile and embedded platforms. 

The standard convolution is defined as
\begin{equation}
(W * y)_{i,\;j\;} = \sum_{k=1}^{K} \sum_{l=1}^{L}  \sum_{m=1}^{M} W_{k, l, m} \; . \; y_{i+k,\;j+l,m} 
\end{equation}
where $W$ is the weight kernel, $y$ is the input feature map, $i, j, m$ are the height, width and channel dimension, and $*$ denotes the standard convolution operation. 

The point-wise $*_{PW}$ , depth-wise $*_{DW}$ and depth separable convolution $DSC(W^p, W^d, y)$ are defined as follows
\begin{eqnarray}
(W *_{PW} y)_{i,\;j\;} = \sum_{m=1}^{M} W_{k, l, m} \; \odot \; y_{i+k,\;j+l,m} \\
(W *_{DW} y)_{i,\;j\;} = \sum_{k=1}^{K} \sum_{l=1}^{L} W_{k, l} \; \odot \; y_{i+k,\;j+l}  \quad \forall \; m \\
DSC(W^p, W^d, y)_{i, \; j} = (W^p *_{PW} (W^d *_{DW} y)_{i,\;j\;})_{i,\;j\;}
\end{eqnarray}
where $W^p$ and $W^d$ represents the different weights for point-wise and depth-wise convolution, and $\odot$ represents element-wise multiplication.

The depth seperable convolution drastically reduces the number of parameters compared to regular convolution as 

\begin{eqnarray}
P_{conv}\;\;\;=\;d_k\; \times \;d_k\; \times \;M\; \times \;N \notag\\
P_{DSC\;}=\;M\;\times\;(d_k\;\times\;d_k)\;+\;M\;\times\;N
\end{eqnarray}
where M denotes number of input channels, N denotes number of output channels and $d_k$ represents kernel size. For e.g., if we have 48 input channels (M), 96 output channels (N) and kernel of size 3 ($d_k$) this would result in 41,472 parameters for regular convolution compared to only 5040 parameters for depth-wise separable convolution, meaning 87.84\% reduction in parameters.

\subsection{Proposed Method}
\begin{figure}[!htb]
	\centering
	{\includegraphics[width=\linewidth]{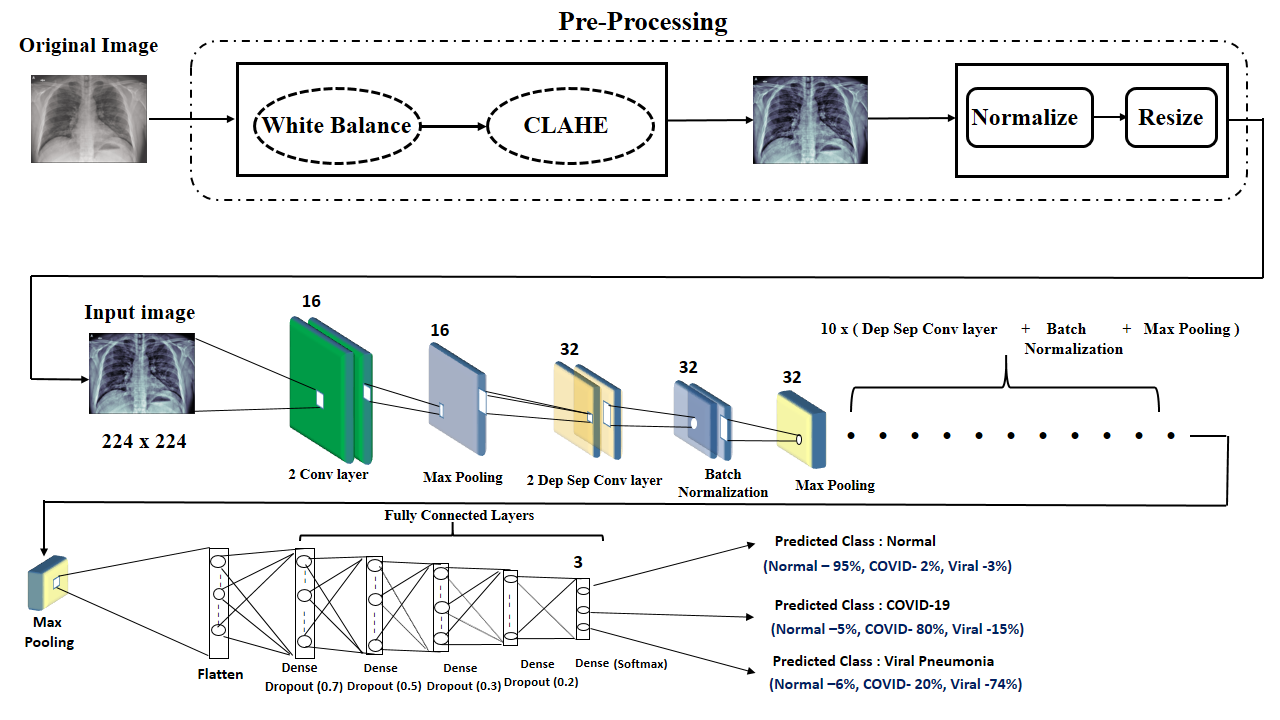} }
	\caption{Training methodology of the proposed neural network solution (COVIDLite)}
	\label{fig_2}
\end{figure}

The methodology proposed in this study for detection of COVID-19 pneumonia from CXR images is depicted in Fig. \ref{fig_2}. 

As shown in Fig. \ref{fig_2}, the overall methodology for training and evaluating the proposed method (COVIDLite) are described in the following steps -

\textit{Step 1:}  Collected the labeled image dataset from different sources and imported them for further pre-processing.

\textit{Step 2:} In this step image pre-processing performed using white balance followed by Contrast Limited Adaptive Histogram Equalization (CLAHE) technique. The White balance and CLAHE method are described below. 

\textbf{White Balance: } White Balance is the image processing operation applied to adjust proper color fidelity in a digital image. Due to low lighting conditions in medical images, some of the parts of the image appeared dark and the image capturing equipment does not detect light precisely as the human eye does. Due to this, image processing or correction help to ensure that the final image represents the colors of the natural image. The objective of this operation is to enhance the visibility of the image so that DCNNs could extract useful features from the image. The white balance algorithm adjusts the colors of the active layers of the image by stretching red, green, and blue channels independently. For doing this, pixel colors discarded, which are at the end of the three channels and are used by only 0.05\% of the pixels in the image, while stretching is performed for the remaining color range. After this operation, pixel colors infrequently present at the end of the channel could not negatively influence the upper and lower bound values while stretching ~\cite{whitebal2020} ~\cite{whitebal2020gdal}. In this solution, we have implemented a white balance algorithm in python language using NumPy and OpenCV library. 

The steps of the White balance algorithm can be summarized as
\begin{eqnarray}
M_i = P_{0.05}(C) \notag \\
M_a = P_{100-0.05}(C) \notag  \\
C_{upd} = \text{Clip}\bigg( \frac{C-M_i)*255}{M_a-M_i}, 0, 255 \bigg) 
\end{eqnarray}

where $P_i(C)$ represents the taking the $i^{\text{th}}$ percentile of channel $C$, and $\text{Clip(., min, max)}$ operation depicts performing saturation operation within min and max values. $C$, $C_{upd}$ denotes the input and updated channels pixel values after the operation respectively. 

\begin{figure}[!htb]
	\centering
	{\includegraphics[width=\linewidth]{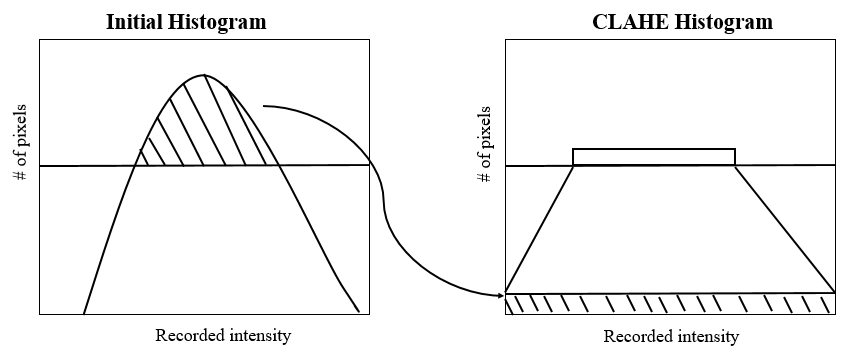} }
	\caption{Working principle of CLAHE showing original and clipped histogram}
	\label{fig_3}
\end{figure}

\textbf{CLAHE:} is an effective contrast enhancement method that effectively increases the contrast of the image. CLAHE is an improved version of the adaptive histogram equation (AHE). Histogram equalization is the simple method for enhancing the contrast of the image by spreading out the intensity range of the image or stretching out the most frequent intensity value of the image. Stretching the intensity values changes the natural brightness of the input image and introduces some undesirable noise in the image ~\cite{pizer1987adaptive}. In AHE, the input image split into several small images, also known as tiles. In this method, the histogram of each tile computed, which corresponds to different sections of the image and uses them to derive intensity remapping function for each tile. This method introduces noise in the image due to over amplification. CLAHE works precisely the same as AHE, but it clips the histogram at specific values for limiting the amplification before computing the cumulative distributive function. The overamplified part of the histogram is further redistributed over the histogram as shown in Fig. \ref{fig_3} ~\cite{pizer1987adaptive}. In one of the previous studies ~\cite{pizer1990contrast}, CLAHE showed exceptional results in enhancing chest CT images and considered useful in examining a wide variety of medical images. The computation of CLAHE ~\cite{wong2014comparative} is performed as 

\begin{equation}
 p\;\;=\;(p_{max}\;-\;p_{min})\;\ast\;P(f)\;+\;p_{min}
\end{equation}
where, $p$ represents pixel value after applying CLAHE, $p_{max} $ , $p_{min} $ represents maximum and minimum pixel value of an image respectively and $P(f)$ represents cumulative probability distribution function.

\begin{figure}[!htb]
	\centering
	{\includegraphics[width=\linewidth]{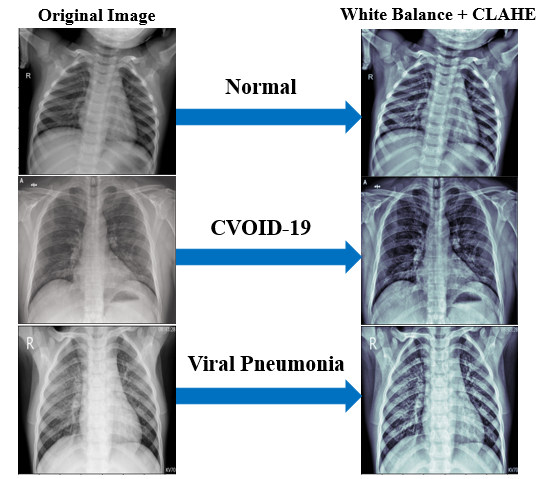} }
	\caption{Pre-processed images after applying white balance followed by CLAHE}
	\label{fig_4}
\end{figure}

The preprocessed images are presented in Fig. \ref{fig_4}, which depicts that white balance and CLAHE have significantly enhanced the clarity of the image and better represented the medical CXR images, which otherwise were not well represented in the original images. 

\textit{Step 3:} This step performs image normalization followed by the resizing of the image. Normalization is an important step that ensures that each pixel in the image has a uniform distribution. It helps in faster convergence while training a deep neural network. In this step, the image pixels are normalized by 255, which scales the pixel values between (0,1). Further, the image is re-sized to $224 \times 224$ pixels before feeding the image to the input layer of DCNN.

\textit{Step 4:} The dataset is split in the ratio of 80\%-20\% for training and testing respectively. 

\textit{Step 5:} The proposed DSC network comprising of 2 convolution layers and 12 depth-wise separable convolution layers is trained. Each convolution layer followed by a max-pooling layer, whereas the depth-wise convolution layer followed by batch normalization and max-pooling layer. In the end fully connected dense layers followed by softmax layer and output layer.

\textit{Step 6:} K-fold cross-validation of DCNN performed with 100 epochs for each fold. The training set is divided into k folds, and the network trained on k-1 parts while validation performed on the remaining 1 part. This step is repeated k times (5 in our case), and the network performance reported by taking the average of k fold predictions.

\textit{Step 7:} Test the network on test data by performing network evaluation using different performance metrics such as plotting confusion matrix, calculating accuracy, precision, sensitivity, specificity, F1-score, and average area under the curve (AUC) for each class type. Furthermore, the t-SNE of the images and the saliency maps of the images are performed for interpretability of the NN.

\subsection{Architecture of proposed COVIDLite}
The proposed network of COVIDLite, comprises of a combination of two separate convolution blocks. The input of the network is a $224\times 224\times 3$ channel CXR image, followed by 18 convolution layers. The size of the kernel used in the network chosen to be small, i.e. $3 \times 3$. The kernel size of $3 \times 3$ also proposed in XceptionNet architecture, ~\cite{chollet2017xception}, which resulted in higher accuracy than InceptionV3 network ~\cite{szegedy2016rethinking} when tested on large image classification dataset, ImageNet dataset  ~\cite{deng2009imagenet}, comprising 350 million images and 17,000 classes. In the first block, two convolution layers followed by a max-pooling layer of size $2 \times 2$. In the second convolution block, two convolution layers are followed by batch normalization, further followed by a max-pooling layer of size $2 \times 2$. The problem of traditional deep neural networks with high learning rate results in the gradient that may explode or vanish or may stuck in sub-optimal local minima ~\cite{ioffe2015batch} . One of the significant problems with the deep neural net is the problem of overfitting. We have solved the problem of overfitting by employing dropout layers proposed in ~\cite{srivastava2014dropout} . We have solved the problem of overfitting by employing dropout layers. We have employed dropout layers after sixth separable convolution layer with dropout ratio of 0.2, which further increased in the range of (0.7 to 0.2) is fully connected dense layers. The detailed architecture of the deep CNN model used in the proposed method shown in Tab. \ref{Tab1}.

{\renewcommand{\arraystretch}{0.8}
\begin{table}[!htb]
\centering
\setlength{\tabcolsep}{4pt}
\begin{tabular}{ |c|c|c|c|c|c|  } 
 \hline
\textbf{Layer Name} & \textbf{O/P shape} & \textbf{Parameters} & \textbf{Kernel Size}& \textbf{Dropout} & \textbf{Filters} \\
 \hline
Input & (224,224,3) & 0 & - & 0 & - \\
Conv2D x 2 & (224,224,16) & 2768 & 3 x 3 & 0 & 4 \\
Maxpool2D & (112,112,16) & 0 & - & 0 & -\\
Separable Conv2D x 2 & (112,112,32) & 2032 & 3 x 3 & 0 & 32\\
Batch Norm. & (112,112,32) & 128 & - & 0 & - \\
Maxpool2D & (56,56,32) & 0 & - & 0 & -\\
Separable Conv2D x 2 & (56,56,64) & 7136 & 3 x 3 & 0 & 64\\
Batch Norm. & (56,56,64) & 256 & - & 0 & -\\
Maxpool2D & (28,28,64) & 0 & - & 0.2 & - \\
Separable Conv2D x 2 & (28,28,128) & 26560 & 3 x 3 & 0 & 128\\
Batch Norm. & (28,28,128) & 512 & - & 0 &-\\
Maxpool2D & (14,14,128) & 0 & - & 0.2 & -\\
Separable Conv2D x 2 & (14,14,256) & 102272 & 3 x 3 & 0 & 256\\
Batch Norm. & (14,14,256) & 1024 & - & 0 & -\\
Maxpool2D & (7,7,256) & 0 & - & 0.2 & -\\
Separable Conv2D x 2 & (7,7,256) & 136192 & 3 x 3 & 0 & 256\\
Batch Norm. & (3,3,256) & 1024 & - & 0 & -\\
Maxpool2D & (3,3,256) & 0 & - & 0.2 & -\\
Separable Conv2D x 2 & (3,3,512) & 401152 & 3 x 3 & 0 & 512\\
Batch Norm. & (3,3,512) & 2048 & - & 0 & -\\
Maxpool2D & (1,1,512) & 0 & - & 0.2 & -\\
FC1 (ReLU) & (512) & 262656 & - & 0.7 & 512\\
FC2 (ReLU) & (128) & 65664 & - & 0.5 & 128\\
FC3 (ReLU) & (64) & 8256 & - & 0.3 & 64\\
FC4 (ReLU) & (32) & 2080 & - & 0.2 & 32\\
FC5 (Softmax) & (3) & 33 & - & 0 & 3\\
\hline
\end{tabular} 
\caption{{Detailed neural network architecture of the proposed method (COVIDLite)} } 
\label{Tab1}
\end{table}
}

\subsubsection{Activation function}  The selection of activation function plays a key role in convergence of the deep neural network. In our proposed solution, we have used ReLU activation function in hidden layers of  the network. ReLU outputs 0 when input value x {\textless} 0, whereas outputs linear function when $x \geq 0 $. In the last layer of the deep CNN model,  softmax function is used for classification. Softmax maps the output of the last layer of the network to the normalized probability distribution over the predicted output classes. Softmax function is defined as 
\begin{equation}
S(X_i)\quad = \frac{\exp(X_i)}{{\displaystyle\sum_{j=0}^{k}}\exp(X_j)} \quad i=1,2,\ldots K 
\end{equation}
where $K$ defines the number of classes to be predicted.

\subsubsection{Loss function} 
As we have multi-class classification problem and the labels are mutually exclusive i.e., each sample belongs to only one class at a time so we have used sparse categorical cross entropy loss function. It is a variant of cross entropy loss function. The basic difference between two loss function is in terms of practical implementation. In case of sparse categorical cross entropy loss function integer labels are used for calculating loss against ground truth unlike one-hot encoded vectors in case of categorical cross entropy loss function which makes them memory efficient by saving lots of computation in terms of log operation and results in faster execution. The sparse categorical cross entropy loss function is defined as    
\begin{equation}
J(\theta) = -\sum_{i=1}^{M} y_i \log \big(\hat{y}_i (\theta) \big)
\end{equation}
where, $\theta$ represents model parameters, $y_i$ denotes true labels in integers, $\hat{y}_i(\theta)$ denotes predicted value by model and $M$ denotes the total number of classes

\section{Dataset and Performance Measures}

\subsection{Dataset}
\begin{figure}[!htb]
	\centering
	{\includegraphics[width=\linewidth]{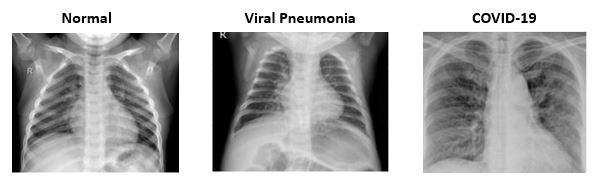} }
	\caption{{Sample chext X-ray images of Normal, Viral Pneumonia and COVID-19 cases}}
	\label{fig_5}
\end{figure}

The dataset used in this study consists of 1823 images of an annotated poster anterior(PA) view of CXR images. Labeled Optical Coherence Tomography (OCT) and CXR Images ~\cite{kermany2018large}  used for viral pneumonia and non-pneumonia or normal cases, whereas three different datasets  ~\cite{cohen2020data,4kcm-m312-20,dsheetieee} used for viral pneumonia and non-pneumonia or normal cases, whereas three different datasets used for COVID-19 cases. The dataset consists of 536 images of COVID-19, 619 images of viral pneumonia, and 668 images of normal cases. The age range of COVID-19 cases in the dataset is 18-75 years{\textemdash}the detailed specification of images used in the dataset depicted in Tab.~\ref{Tab2}. As shown in Tab.~\ref{Tab2}, COVID-19 images have a considerable variation in height and width with maximum height and maximum width compared to other class of images. Figure~\ref{fig_5} shows the sample images of normal, viral pneumonia, and COVID-19 cases. As per Fig.~\ref{fig_5}, the normal CXR image depicts clear lungs without any area of abnormal opacification or pattern in the image, viral pneumonia (middle) manifests with a more diffuse "interstitial" pattern in both lungs and COVID-19 (extreme right) depicts ground-glass opacification and consolidation in the right upper lobe and left lower lobe. For training deep CNN model, the dataset divided into a ratio of 80-20, where 80\% of the dataset used for training the model while 20\% used for testing the model. The distribution of images in the training and test set shown in Tab.~\ref{Tab3}. For the analysis of CXR images, all CXR images initially screened for quality control by filtering all low quality or unreadable scans for building efficient deep learning models.

{\renewcommand{\arraystretch}{1.0}
\begin{table}[h!]
\centering
\setlength{\tabcolsep}{8pt}
\begin{tabular}{ |c|c|c|c|c| } 
 \hline
\textbf{Image Class} & \textbf{Min. width} & \textbf{Max. width} & \textbf{Min. height}& \textbf{Max. height} \\
 \hline
Normal & 1040 & 2628 & 650 & 2628\\
COVID-19 & 240 & 4095 & 237 & 4095\\
Viral Pneumonia & 384 & 2304 & 127 & 2304\\
\hline
\end{tabular} 
\caption{{Detailed specification of images in the dataset} } 
\label{Tab2}
\end{table}
}

{\renewcommand{\arraystretch}{1.0}
\begin{table}[h!]
\centering
\setlength{\tabcolsep}{8pt}
\begin{tabular}{ |c|c|c|} 
 \hline
\textbf{Image Class} & \textbf{Training Set} & \textbf{Test Set} \\
 \hline
Normal &  534 & 134\\
COVID-19 &  429 &  107\\
Viral pneumonia &  495 &  124\\
\hline
Total & 1458 & 365\\
\hline
\end{tabular} 
\caption{Distribution of images in training and test set for 3-class problem } 
\label{Tab3}
\end{table}
}

\begin{figure}[!htb]
	\centering
	{\includegraphics[width=0.7\linewidth]{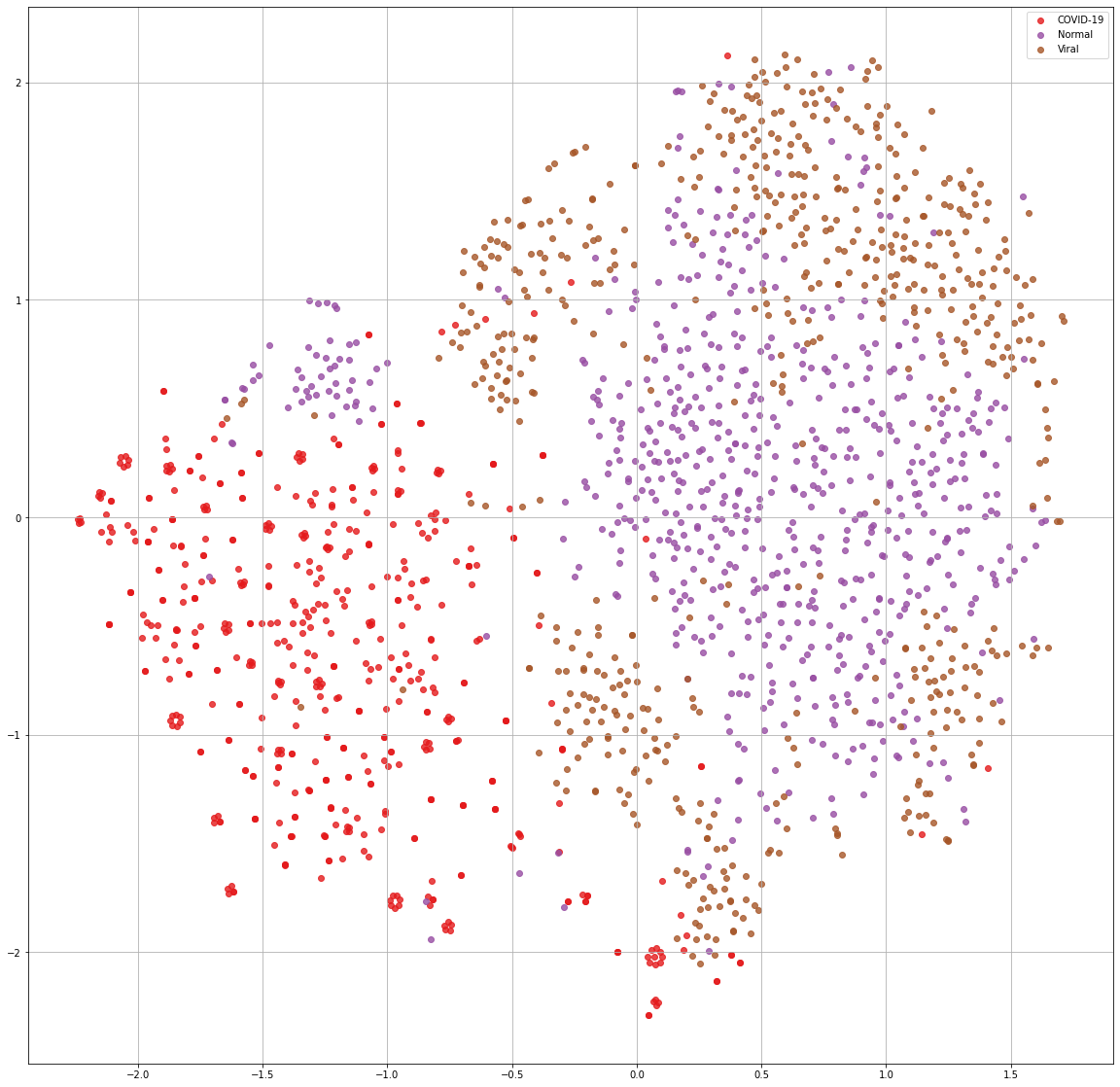} }
	\caption{{t-SNE plot depicting separability among different classes of input image data in the dataset}}
	\label{fig_6}
\end{figure}

Figure~\ref{fig_6} demonstrates the separability among different classes i.e., Normal, COVID-19 and viral pneumonia using t-SNE plot on the normalized input image data.

\subsection{Performance Measures}
For evaluating the performance of the proposed solution, several performance metrics are chosen, i.e., sensitivity, specificity, precision, F1-score, accuracy, an average area under the curve (AUC), and cohen's kappa score. First, confusion matrix plotted based on the proposed model's predictions to assess the number of instances correctly classified and misclassified by the model. The confusion matrix consists of true positives (TP), true negatives (TN), false positives (FP) and false negatives (FN). True positives (TP) are the number of disease cases correctly classified by the model, true negatives (TN) are actual number of normal cases correctly classified as normal. False positives (FP) are the actual number of normal cases misclassified as a disease. False negatives (FN) are the number of disease cases misclassified as normal. The definitions of sensitivity, specificity, precision, F1-score, and accuracy are as follows -

\begin{eqnarray}
\text{Sensitvitiy}\;=\;\frac{TP}{TP\;+\;FN} \notag \\
\text{Specificity} \;=\;\frac{TN}{TN\;+\;FP} \notag \\
\text{Precision} \;=\;\frac{TP}{TP\;+\;FP} \notag \\
\text{F1}\;=\;2\;\times\;\frac{\text{Precision}\;\times\;\text{Recall}}{\text{Precision}\;+\;\text{Recall}} \notag \\
\text{Accuracy}\;=\;\frac{TP\;+\;TN}{TP\;+\;TN\;+\;FP\;+\;FN} \notag \\
\text{Cohen's kappa score} = \frac{p_o\;-\;p_e}{1\;-\;p_e}
\end{eqnarray}

where, $p_o $ represents  relative observed agreement between ground truth and the model, $p_e $ represents expected agreement between ground truth and the model which determines superiority of the model in comparison to simple random guess based on distribution of each class.

The receiver operating characteristic (ROC) curve is the graph plotted between True Positive Rate (TPR) and False Positive Rate (FPR). TPR and FPR are defined as 

\begin{eqnarray}
\text{True Positive Rate (TPR)} \;=\; \frac{TP}{TP\;+\;FN} \notag \\
\text{False Positive Rate (FPR)} \;=\; \frac{FP}{FP\;+\;TN}
\end{eqnarray}

Average Area Under the ROC curve (AUC) measures the two-dimensional area under the ROC curve from point (0,0) to (1,1). For multi-class problem, AUC is computed as an extension of multi-class AUC as proposed in \unskip~\cite{tang2011towards}.

Furthermore, Grad-CAM maps, and Locally model-agnostic explanations visualization methods are used for interpretability of the trained neural network and gain insights of the featuers the neural network is basing its decisions on.
    
\section{Experimental results}

\subsection{Experimental settings}
In this paper, we have keras library with tensorflow as backend is used for building the deep CNN model. The Linux operating system is used with Intel Xeon CPU E3-1225, 3.3 GHz, and 8 GB RAM. For evaluating the proposed method, the train-cross-validation-test scheme is used in which training of deep CNN model performed on the training set, 5-fold cross-validation used for tuning the hyper-parameters of the model and finally test set used for evaluating the proposed method. The deep CNN model trained using a mini-batch size of 8. The Adam optimizer used with weight decay in terms of initial learning rate / total number of epochs. The initial learning rate to be taken as 0.001 and the maximum number of epochs to train the network was 100.

\subsection{Results}
In this section, we have demonstrated the performance of the proposed method using different performance metrics. Table~\ref{Tab4} showed the summary of the 5-fold cross-validation accuracy of the proposed COVIDLite method. As shown in Table~\ref{Tab4}, the average 5-fold cross-validation accuracy in the case of binary classification reported as 98.02\% slightly higher than multi-class classification accuracy of 97.12\%. Table~\ref{Tab5} demonstrates the class-wise performance of the proposed method for multi-class classification problems. The macro and weighted average results calculated using the classification report function of the sklearn library in python. The macro and weighted average F1-score of the model is 96\%. As shown in Table~\ref{Tab5}, the model is highly specific and highly sensitive in predicting COVID-19 cases.

{\renewcommand{\arraystretch}{1.0}
\begin{table}[h!]
\centering
\setlength{\tabcolsep}{8pt}
\begin{tabular}{ |c|c|c|} 
 \hline
\textbf{Number of Folds} & \textbf{Accuracy (2-class)} & \textbf{Accuracy (3-class)} \\
 \hline
Fold 1 &   98.26\% &  94.52\%\\
Fold 2 &  99.48\% &  96.23\%\\
Fold 3 &  98.96\% &  96.58\%\\
Fold 4 &  100.00\% &  98.97\%\\
Fold 5 &  92.71\% &  99.31\%\\
Average & 98.02\% ($\pm$ 2.69\%) &  97.12\% ($\pm$ 1.79\%)\\
\hline
\end{tabular} 
\caption{Cross -validation performance of the proposed method for 2-class vs 3-class problem} 
\label{Tab4}
\end{table}
}

{\renewcommand{\arraystretch}{0.9}
	\begin{table}[h!]
		\centering
		\setlength{\tabcolsep}{6pt}
		\begin{tabular}{ |c|c|c|c|} 
			\hline
			Parameters & Normal & COVID-19 & Viral Pneumonia \\
			\hline
			Precision &  96\% &  98\% &  95\% \\
			Sensitivity &  97\% &  97\% &  95\% \\
			Specificity &  99.11\% &  99.18\% &  94.58\% \\
			F1-Score &  97\% &  98\% &  95\% \\
			Class error (95\% CI) &  (0.00103, 0.05866) & (0.00324, 0.05930)  &  (0.01061, 0.08614) \\
			AUC &  0.99 &  1.0 &  1.0 \\
			\hline
		\end{tabular} 
		\caption{Class-wise performance of the proposed method for 3-class problem} 
		\label{Tab5}
	\end{table}
}

In Table~\ref{Tab6}, we demonstrated the performance summary of the proposed COVIDLite model with and without pre-processing step, i.e., White balance followed by CLAHE for both 2-class and 3-class scenario. As per the table proposed method with pre-processing resulted in improved performance of approx 1\% in accuracy, precision and F1-score, and approx 2\% improvement observed in cohen's kappa score for the 3-class problem. In comparison, for binary classification, approx 1\% improvement observed in accuracy, precision, specificity, and approx 2\% increase observed in terms of cohen's kappa score. In both the binary and multi-class problem significant increase observed in performance measures after applying White balance followed by CLAHE as well as 95\% confidence intervals are narrower in comparison to method with no pre-processing step i.e., White balance followed by CLAHE.

{\renewcommand{\arraystretch}{1.0}
	\begin{table}[h!]
		\centering
		\setlength{\tabcolsep}{4pt}
		\begin{tabular}{ |c|c|c|c|c|} 
			\hline
			Parameters & \multicolumn{1}{|p{3cm}|}{\centering 2-class \\ (no WB+CLAHE)} & \multicolumn{1}{|p{2cm}|}{\centering 2-class \\ (proposed)} & \multicolumn{1}{|p{3cm}|}{\centering 3-class \\ (no WB+CLAHE)} &\multicolumn{1}{|p{2cm}|}{\centering 3-class \\ (proposed)} \\
			\hline
			Accuracy &  98.75 &  99.58 &  95.34 &  96.43\\
			Precision &  99.00 &  100.00 &  96.00 &  97.00\\
			Sensitivity &  99.00 &  99.58 &  96.00 &  96.00\\
			Specificity &  98.25 &  99.34 &  97.79 &  97.89\\
			F1-Score &  99.00 &  99.79 &  95.00 &  96.00\\
			Class error (95\% CI) &  (0.00155, 0.02643) &  (0.0067, 0.02103) & (0.02709, 0.07152) &  (0.0165, 0.0546)\\
			Kappa &  0.9748 &  0.9916 &  0.9299 &  0.9463\\
			AUC &  1.0 &  1.0 &  0.99 &  0.99\\
			\hline
		\end{tabular} 
		\caption{Performance summary of proposed method with and without WB+CLAHE for 2-class vs 3-class problem} 
		\label{Tab6}
	\end{table}
}

\begin{figure}[!htb]
	\centering
	{\includegraphics[width=1.0\linewidth]{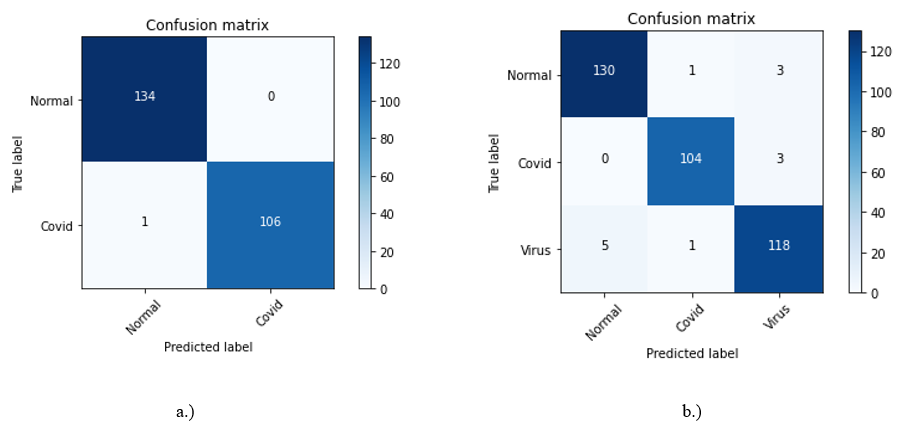} }
	\caption{{Confusion matrix of proposed method for a) 2-class b) 3-class}}
	\label{fig_7}
\end{figure}

Figure ~\ref{fig_7} demonstrates the confusion matrix based on the classification result of the model's prediction for binary class and multi-class problems. In a multi-class problem, the model predicted 4 False positives out of which one belongs to COVID-19, and three belongs to viral pneumonia. Due to overlapping patterns exist between COVID-19 and viral pneumonia, the model predicted 3 COVID-19 cases as viral pneumonia, whereas one viral pneumonia case predicted as COVID-19. Further, in the case of viral pneumonia, the model predicted 5 False negatives in which five viral pneumonia cases predicted as normal while 1 case predicted as COVID-19. In the case of COVID-19, the model predicted three false negatives, and all the three COVID cases predicted as viral pneumonia. In binary classification, the model has predicted only 1 False negative, which showed that the model is highly sensitive, precise, and specific in predicting COVID cases.

\begin{figure}[!htb]
	\centering
	{\includegraphics[width=1.2\linewidth]{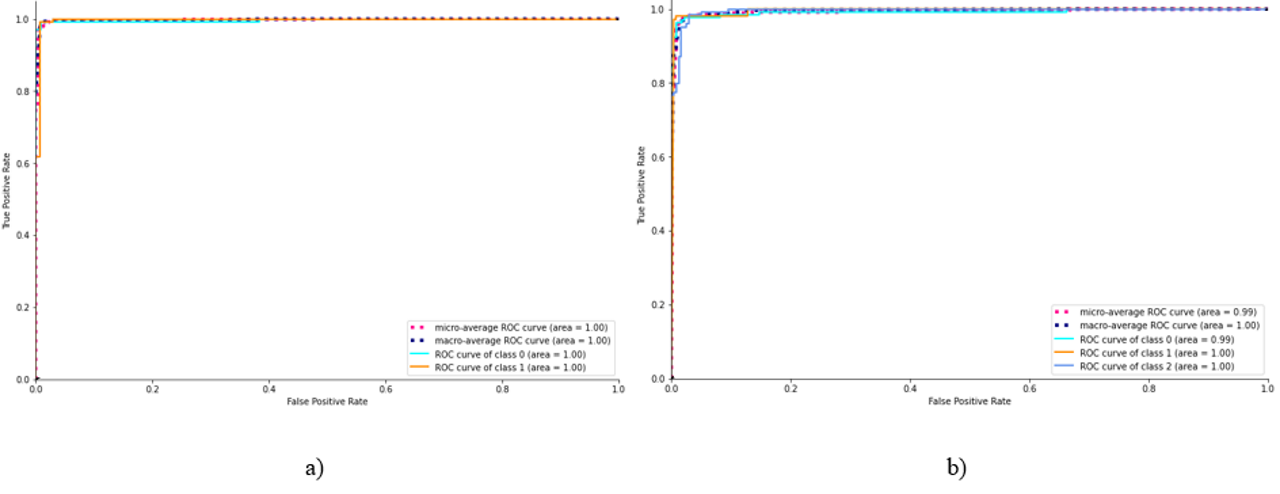} }
	\caption{{ROC curve of proposed method COVIDLite a) 2-class b) 3-class problem}}
	\label{fig_8}
\end{figure}

In Fig.~\ref{fig_8}, we presented the ROC curve for both binary classification and multi-class classification problem. In multi-class problem the average area under the curve (AUC) for class 0 i.e., normal cases is 0.99, for class 1 i.e., COVID-19 class is 1.0 and for class 2 i.e., viral pneumonia is 0.99. However, in case of binary classification the average area under the curve (AUC) for both COVID-19 and normal class is 1.0.

Further, we have shown the LIME ~\cite{ribeiro2016should}  maps, saliency maps and Grad-CAM ~\cite{selvaraju2017grad}  visualization of misclassified instances of COVIDLite without pre-processing for viral and COVID-19 patients in Figure 9 and correctly classified instances of COVIDLite with pre-processing (White Balance + CLAHE) in Figure 10. For detecting COVID-19 from CXR images the most common findings include lung consolidation and ground glass opacities ~\cite{jacobi2020portable} . Apart from that, COVID-19 and other viral pneumonia typically cause lung opacities in more than one lobe ~\cite{jacobi2020portable} . Earlier researchers noted that patients suffering from COVID-19 include air-space disease which involve  bilateral lower lung distribution ~\cite{wong2020frequency} .

\begin{figure}[!htb]
	\centering
	{\includegraphics[width=1.0\linewidth]{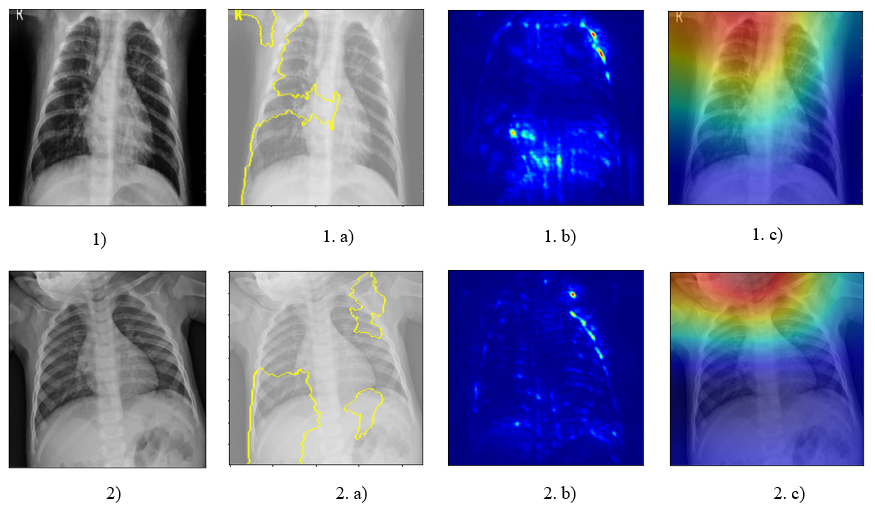} }
	\caption{Visualization of incorrect predictions of proposed method without (White balance + CLAHE). Images 1 and 2 show the original CXRs of viral and COVID-19, images 1.a) and 1.b) show the LIME maps, images 1.b) and 2.b) show the saliency maps and images 1.c) and 2.c) show the grad-CAM heatmap of viral and COVID-19 patients respectively.}
	\label{fig_9}
\end{figure}

\begin{figure}[!htb]
	\centering
	{\includegraphics[width=1.0\linewidth]{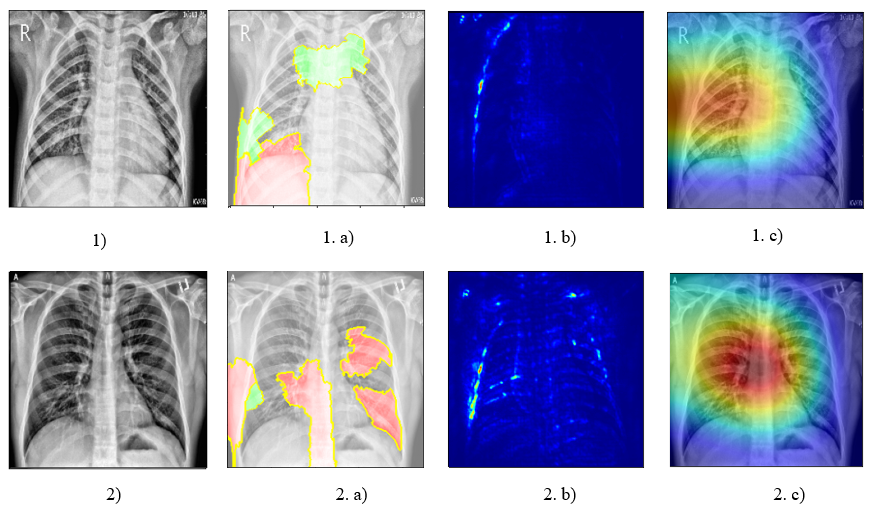} }
	\caption{Visualization of correct predictions of proposed method with (White balance + CLAHE). Images 1 and 2 show the original CXRs of viral and COVID-19, images 1.a) and 1.b) show the LIME maps, images 1.b) and 2.b) show the saliency maps and images 1.c) and 2.c) show the grad-CAM heatmap of viral and COVID-19 patients respectively.}
	\label{fig_10}
\end{figure}

As shown in Figure~\ref{fig_9}, without pre-processing model considered noisy features for prediction of viral and COVID-19 induced pneumonia and wrongly predicted both the cases as normal. Figure~\ref{fig_9} 1a) and Fig.~\ref{fig_9} 2a) demonstrates the LIME map of the model's prediction outlining localized regions in CXR images while Fig.\ref{fig_9} 1c) and Fig.~\ref{fig_9} 2c) demonstrate the Grad-CAM heatmap highlighting the upper region of the CXR. The saliency map highlighting the lower and upper right corner of CXR in Fig.\ref{fig_9} 1b) and highlighting the upper right corner of CXR in Fig.\ref{fig_9} 2b) instead of focusing on ground glass opacity and pneumonia involving the lower lung zones bilaterally on CXR in COVID-19 case and on diffuse interstitial pattern showing lung inflammation in right lung in viral pneumonia case. Thus, the model without pre-processing was unable to detect core features for classifying viral and COVID-19 pneumonia resulted in incorrect prediction whereas model with pre-processing, i.e., White balance followed by CLAHE correctly highlighted region affected with pneumonia, i.e., left lung inflammation as shown in Figure Fig.~\ref{fig_10} 1a), Fig.~ \ref{fig_10} 1b) and Fig.~\ref{fig_10} 1c) and accurately highlighted peripheral left mid to lower lung opacities and ground glass opacity (generally visible in chest CT) in COVID-19 case as shown in Fig.\ref{fig_10} 2a), Fig.~\ref{fig_10} 2b) and Fig.\ref{fig_10}  2c) respectively.

\section{Discussion}
This section illustrates the comparison of proposed method with recent state-of-the-art methods applied for detection of COVID-19 using radiology images i.e., CXR and chest CT. In most of the studies researchers used Dr Cohen's image collection dataset ~\cite{cohen2020data}. This was the first open source dataset available for researchers to extract useful patterns from CXR and chest CT images for detection of COVID-19. ~\cite{apostolopoulos2020covid} proposed transfer learning based pre-trained network VGG19 for detecting COVID-19 from CXR images. In this study researchers have used 224 CXR images of COVID-19, 700 images of viral pneumonia and 504 images of normal or healthy cases. Their model achieved an accuracy of 93.48\% for 3-class problem. ~\cite{wang2020covid} used light-weight deep learning architecture named as COVID-Net. They have used 13,975 CXR images across 13,870 patients and achieved a test accuracy of 93.3\%. In this study researchers have employed projection-expansion-projection-extension (PEPX) design pattern comprising 1 x1 convolution at beginning and end of architecture while depth-wise convolutions at the middle. Xu et al. ~\cite{xu2020others}  proposed combination of pre-trained network ResNet and Location Attention method. They have achieved an accuracy of 86.7\% with 219 COVID-19, 224 viral pneumonia and 175 normal chest CT images. ~\cite{ozturk2020automated}  developed deep CNN model named DarkCovidNet for detection of COVID-19 from CXR images. They have used DarkNet-19 ~\cite{redmon2017yolo9000} architecture for building the model and achieved an accuracy of 87.02\% for multi-class and 98.08\% for 2-class problem. ~\cite{abbas2020classification}  proposed class decomposition mechanism named DeTraC for investigating class boundaries with ResNet18 pre-trained network for detection of COVID-19 from 196 CXR images. Further they have used PCA for feature space dimensionality reduction resulted in achieving higher accuracy of 95.12\%. ~\cite{luz2020towards} proposed two variants of EfficientNet deep learning model i.e., hierarchical and Flat. In hierarchical model researchers used one classifier at root node of the tree for classifying Normal and viral pneumonia cases and the second classifier at higher level to further classify COVID-19 and viral Pneumonia cases. This approach resulted in lower accuracy of 93.50\% in comparison to Flat model which attain overall test accuracy of 93.9\%. Both flat efficient net B0 and hierarchical efficientnetB3 are larger in terms of size and total parameters having total parameters 5,330,564 (21 MB) and 12,320,528 (48 MB) respectively. 

In case of binary classification problem, researchers have used different combination of transfer learning approaches. ~\cite{narin2020automatic}  employed three pre-trained networks (ResNet50, InceptionV3, and InceptionResNetV2) in which ResNet50 outperformed other variants by achieving a higher accuracy of 98\%. In this study, researchers used 50 CXR images of COVID-19 positive patients and 50 images of COVID-19 negative images. ~\cite{sethy2020detection}  used ResNet50 in combination with support vector machines (SVM) for extracting deep features. However, in this study researchers have used a very small dataset of 50 CXR images comprising 25 COVID-19 images and 25 COVID-19 negative images. ~\cite{song2020deep}  employed a combination of Attention and Feature Pyramid Network (FPN) to extract top K features of chest CT images using deep ResNet pre-trained network. In this study, they have achieved an accuracy of 86\% with 777 COVID-19 images and 708 healthy images. ~\cite{wang2020deep}  proposed a modified Inception network (M-Inception) to detect COVID-19 from chest CT images with an accuracy of 82.90\%. ~\cite{zheng2020deep}  developed a 3 dimensional CNN model with pre-trained network U-Net using 313 chest CT images of COVID-19 and 229 images of healthy cases. They have reported an accuracy of 90.8\%. ~\cite{rahimzadeh2020modified}  concatenated pre-trained network Xception and ResNet50 V2 for detection of COVID-19 and viral pneumonia using 15085 CXR images. As they have combined two complex architecture for the task the resulting model was computationally intensive with model size of 560 MB which makes it unfavorable in terms of practical implementation for real time predictions. Finally, ~\cite{khan2020coronet} proposed a deep CNN model named CoroNet which uses local receptive field and sharing of weight for better performance. They have attained an overall test accuracy of 95\% for 3-class classification task. Their proposed CNN model was significantly deeper having large number of trainable parameters i.e., 33,915,436 due to which their model size also increased making their method unfavorable for practical implementation in terms of mobile based web app for real time predictions.

In this research, we have developed a deep CNN model for the detection of COVID-19 induced pneumonia using 1823 CXR images. The proposed method named COVIDLite is significantly lighter, with 1019330 total trainable parameters and 8.4 MB in size, making it favorable to be integrated with mobile devices for real-time predictions. Our method has attained an accuracy of 99.58\% for binary classification and 96.43\% for multi-class classification. In this research, we have found that our proposed COVIDLite method outperformed recent state-of-the-art methods employed for the detection of COVID-19 from radiology images in terms of accuracy, as shown in Table~\ref{Tab7}.

{\renewcommand{\arraystretch}{0.7}
\begin{table}[!htb]
\centering
\setlength{\tabcolsep}{4pt}
\begin{tabular}{c|c|c|c}\hline
\textbf{Method} & \textbf{Image Type} & \textbf{Image distribution}  & \textbf{Accuracy (\%)} \\ \hline
VGG19 \cite{apostolopoulos2020covid}   & CXR                 & \begin{tabular}[c]{@{}l@{}}224 COVID-19,\\ 700 Pneumonia,\\ 504 Healthy\end{tabular}                                               & 93.48                                                                   \\ \hline
COVID-Net \cite{wang2020covid}       & CXR                 & \begin{tabular}[c]{@{}l@{}}53 COVID-19(+),\\ 5526 COVID-19(-),\\ 8066 Healthy\end{tabular}                                         & 92.40                                                                   \\ \hline
ResNet + Local Attention \cite{xu2020others}        & CT                  & \begin{tabular}[c]{@{}l@{}}219 COVID-19,\\ 224 Pneumonia,\\ 75 Healthy\end{tabular}                                                & 86.70                                                                   \\ \hline
Dark Covid Net \cite{ozturk2020automated}              & CXR                 & \begin{tabular}[c]{@{}l@{}}125 COVID-19,\\ 500 Pneumonia,\\ 500 No-Findings\end{tabular}                                           & 87.02                                                                   \\ \hline
DeTrac ResNet18  \cite{abbas2020classification}         & CXR                 & \begin{tabular}[c]{@{}l@{}}105 COVID-19,\\ 11 SARS,\\ 80 Normal\end{tabular}                                                       & 95.12                                                                   \\ \hline
Flat-Efficient Net \cite{luz2020towards}         & CXR                 & \begin{tabular}[c]{@{}l@{}}8066 Normal,\\ 5521 Pneumonia,\\ 183 COVID-19\end{tabular}                                              & 93.93                                                                   \\ \hline
Hierarcical Efficient Net B3 \cite{luz2020towards}   & CXR                 & \begin{tabular}[c]{@{}l@{}}8066 Normal,\\ 5521 Pneumonia,\\ 183 COVID-19\end{tabular}                                              & 93.50                                                                   \\ \hline
Deep CNN ResNet-50 \cite{narin2020automatic}      & CXR                 & \begin{tabular}[c]{@{}l@{}}50 COVID-19(+),\\ 50 COVID-19(-)\end{tabular}                                                           & 98.00                                                                   \\ \hline
DRE-Net \cite{song2020deep} & CT                  & \begin{tabular}[c]{@{}l@{}}777 COVID-19(+),\\ 708 Healthy\end{tabular}                                                             & 86.00                                                                   \\ \hline
M-Inception \cite{wang2020deep}  & CT                  & \begin{tabular}[c]{@{}l@{}}195 COVID-19(+), \\ 258 COVID-19(-)\end{tabular}                                                        & 82.90                                                                   \\ \hline
U Net + 3D Deep Network \cite{zheng2020deep}      & CT                  & \begin{tabular}[c]{@{}l@{}}313 COVID-19(+),\\ 229 COVID-19(-)\end{tabular}                                                         & 90.80                                                                   \\ \hline
Xception + ResNet50V2 \cite{rahimzadeh2020modified} & CXR                 & \begin{tabular}[c]{@{}l@{}}180 COVID-19 (+),\\ 6054 Pneumonia,\\ 8851 Normal\end{tabular}                                          & 91.40                                                                   \\ \hline
CoroNet \cite{khan2020coronet} & CXR                 & \begin{tabular}[c]{@{}l@{}}310 Normal,\\ 327 Pneumonia,\\ 284 COVID-19\end{tabular}                                                & 95.00                                                                   \\ \hline
\textbf{Proposed Method}                          & \textbf{CXR}        & \textbf{\begin{tabular}[c]{@{}l@{}}536 COVID-19,\\ 619 Viral pneumonia,\\ 668 Normal\end{tabular}} & \textbf{\begin{tabular}[c]{@{}l@{}}96.43 (3 class)\\ \\  99.58 (2 class)  
 \end{tabular}}  
\\ \hline
\end{tabular}
\caption{Comparison of proposed method with recent state-of-the-art methods for COVID-19 detection using radiology images} 
\label{Tab7}
\end{table}
}
   
\section{Conclusion }
In this paper, we demonstrated an effective deep learning model for detecting COVID-19 from CXR images with higher accuracy. Our proposed method performs both binary and multi-class classification with an accuracy of 99.58\% and 96.43\%, respectively. The proposed method captures low-level feature maps by enhancing the visibility of CXR images with advanced preprocessing techniques, which facilitates recognizing intricate patterns from medical images at a level comparable with experienced radiologists. As future work, we will further enhance our method's performance by including the lateral view of CXR images in our training data, as in some of the cases, frontal view of CXR images does not give a clear idea in diagnosing pneumonia cases. Further, similar proposed preprocessing with the DSCNN method can be extenedd to detect more critical diseases such as lung cancer, fibrosis, tuberculosis, and pnemo-thorax, which in turn can assist the radiologists in the remotest parts of the world, where medical resources are limited.

\bibliographystyle{elsarticle-num}

\bibliography{article}

\begin{thebibliography}{10}
\expandafter\ifx\csname url\endcsname\relax
  \def\url#1{\texttt{#1}}\fi
\expandafter\ifx\csname urlprefix\endcsname\relax\def\urlprefix{URL }\fi
\expandafter\ifx\csname href\endcsname\relax
  \def\href#1#2{#2} \def\path#1{#1}\fi

\bibitem{771142:18601400}
S.~Adhikari, S.~Meng, Y.~Wu, {Epidemiology, causes, clinical manifestation and
  diagnosis, prevention and control of coronavirus disease (COVID-19) during
  the early outbreak period: a scoping review}, {Infect Dis Poverty} 9 (2020)
  29--29.

\bibitem{771142:18601416}
Y.~Yi, P.~N. Lagniton, S.~Ye, E.~Li, R.-H. Xu,
  \href{https://dx.doi.org/10.7150/ijbs.45134}{{COVID-19: what has been learned
  and to be learned about the novel coronavirus disease}}, {International
  Journal of Biological Sciences} 16~(10) (2020) 1753--1766.
\newblock \href {https://doi.org/10.7150/ijbs.45134}
  {\path{doi:10.7150/ijbs.45134}}.
\newline\urlprefix\url{https://dx.doi.org/10.7150/ijbs.45134}

\bibitem{771142:18601395}
B.~Udugama, P.~Kadhiresan, H.~N. Kozlowski, A.~Malekjahani, M.~Osborne,
  V.~Y.~C. Li, H.~Chen, S.~Mubareka, J.~B. Gubbay, W.~C.~W. Chan,
  \href{https://dx.doi.org/10.1021/acsnano.0c02624}{{Diagnosing COVID-19: The
  Disease and Tools for Detection}}, {ACS Nano} 14~(4) (2020) 3822--3835.
\newblock \href {https://doi.org/10.1021/acsnano.0c02624}
  {\path{doi:10.1021/acsnano.0c02624}}.
\newline\urlprefix\url{https://dx.doi.org/10.1021/acsnano.0c02624}

\bibitem{771142:18601417}
J.~P. Kanne, B.~P. Little, J.~H. Chung, B.~M. Elicker, L.~H. Ketai, {Essentials
  for radiologists on COVID-19: an update-radiology scientific expert panel},
  {Radiology} (2020).

\bibitem{771142:18601390}
X.~Xie, Z.~Zhong, W.~Zhao, C.~Zheng, F.~Wang, J.~Liu,
  \href{https://dx.doi.org/10.1148/radiol.2020200343}{{Chest CT for Typical
  2019-nCoV Pneumonia: Relationship to Negative RT-PCR Testing}}, {Radiology}
  (2020) 200343--200343\href {https://doi.org/10.1148/radiol.2020200343}
  {\path{doi:10.1148/radiol.2020200343}}.
\newline\urlprefix\url{https://dx.doi.org/10.1148/radiol.2020200343}

\bibitem{771142:18601418}
W.~Kong, P.~P. Agarwal,
  \href{https://dx.doi.org/10.1148/ryct.2020200028}{{Chest Imaging Appearance
  of COVID-19 Infection}}, {Radiology: Cardiothoracic Imaging} 2~(1) (2020)
  e200028--e200028.
\newblock \href {https://doi.org/10.1148/ryct.2020200028}
  {\path{doi:10.1148/ryct.2020200028}}.
\newline\urlprefix\url{https://dx.doi.org/10.1148/ryct.2020200028}

\bibitem{771142:18601409}
E.~Y.~P. Lee, M.-Y. Ng, P.-L. Khong,
  \href{https://dx.doi.org/10.1016/s1473-3099(20)30134-1}{{COVID-19 pneumonia:
  what has CT taught us?}}, {The Lancet Infectious Diseases} 20~(4) (2020)
  384--385.
\newblock \href {https://doi.org/10.1016/s1473-3099(20)30134-1}
  {\path{doi:10.1016/s1473-3099(20)30134-1}}.
\newline\urlprefix\url{https://dx.doi.org/10.1016/s1473-3099(20)30134-1}

\bibitem{771142:18601407}
H.~Shi, X.~Han, {Radiological findings from 81 patients with COVID-19 pneumonia
  in Wuhan, China: a descriptive study}, {Lancet Infect. Dis}  2020--2020.

\bibitem{771142:18601384}
W.~Zhao, Z.~Zhong, X.~Xie, Q.~Yu, J.~Liu, {Relation between chest CT findings
  and clinical conditions of coronavirus disease (COVID-19) pneumonia: a
  multicenter study}, {Am. J. Roentgenol} 2020  1--6.

\bibitem{771142:18601408}
Y.~Li, L.~Xia, {Coronavirus Disease 2019 (COVID-19): role of chest CT in
  diagnosis and management}, {Am. J. Roentgenol} 2020  1--7.

\bibitem{771142:18601398}
S.~H. Yoon, K.~H. Lee, {Chest radiographic and CT findings of the 2019 novel
  coronavirus disease (COVID-19): analysis of nine patients treated in Korea},
  {Korean J. Radiol} 21~(4) (2020) 494--500.

\bibitem{771142:18601387}
O.~Stephen, M.~Sain, U.~J. Maduh, D.-U. Jeong,
  \href{https://dx.doi.org/10.1155/2019/4180949}{{An Efficient Deep Learning
  Approach to Pneumonia Classification in Healthcare}}, {Journal of Healthcare
  Engineering} 2019 (2019) 1--7.
\newblock \href {https://doi.org/10.1155/2019/4180949}
  {\path{doi:10.1155/2019/4180949}}.
\newline\urlprefix\url{https://dx.doi.org/10.1155/2019/4180949}

\bibitem{771142:18601394}
P.~Chhikara, P.~Singh, P.~Gupta, T.~Bhatia, {Deep Convolutional Neural Network
  with Transfer Learning for Detecting Pneumonia on Chest X-Rays}, in: J.~L.,
  V.~M., P.~V., B.~V. (Eds.), {Advances in Bioinformatics, Multimedia, and
  Electronics Circuits and Signals. Advances in Intelligent Systems and
  Computing}, Vol. 1064, Springer, 2020.

\bibitem{771142:18601388}
Sethy, P.~. Santi, K.~. Behera, .~. Kumar, Pradyumna, {Detection of coronavirus
  Disease (COVID-19) based on Deep Features and Support Vector Machine} (2020).

\bibitem{771142:18601379}
M.~Loey, F.~Smarandache, N.~E.~M. Khalifa,
  \href{https://dx.doi.org/10.3390/sym12040651}{{Within the Lack of Chest
  COVID-19 X-ray Dataset: A Novel Detection Model Based on GAN and Deep
  Transfer Learning}} (2020).
\newblock \href {https://doi.org/10.3390/sym12040651}
  {\path{doi:10.3390/sym12040651}}.
\newline\urlprefix\url{https://dx.doi.org/10.3390/sym12040651}

\bibitem{771142:18601403}
E.~E.~D. Hemdan, M.~A. Shouman, M.~E. Karar,
  \href{arXiv:2003.11055}{{COVIDX-Net: A Framework of Deep Learning Classifiers
  to Diagnose COVID-19 in X-Ray Images}} (2020).
\newline\urlprefix\url{arXiv:2003.11055}

\bibitem{771142:18744830}
S.~Hazra, A.~Santra,
  \href{https://dx.doi.org/10.1109/lsens.2018.2882642}{{Robust Gesture
  Recognition Using Millimetric-Wave Radar System}}, {IEEE Sensors Letters}
  2~(4) (2018) 1--4.
\newblock \href {https://doi.org/10.1109/lsens.2018.2882642}
  {\path{doi:10.1109/lsens.2018.2882642}}.
\newline\urlprefix\url{https://dx.doi.org/10.1109/lsens.2018.2882642}

\bibitem{771142:18744916}
J.~McIntosh, A.~Marzo, M.~Fraser, C.~Phillips, { Echoflex: Hand gesture
  recognition using ultrasound imaging}, in: Proceedings of the 2017 CHI
  Conference on Human Factors in Computing Systems, 2017, pp. 1215--1247.

\bibitem{771142:18744941}
Picone, J.W, {Signal modeling techniques in speech recognition}, in:
  Proceedings of the IEEE, 1993, pp. 1215--1247.

\bibitem{771142:18601397}
F.~Chollet, {Xception: Deep learning with Depthwise Separable Convolutions},
  in: {Proceedings of the IEEE Conference on Computer Vision and Pattern
  Recognition}, 2017, pp. 1251--1258.

\bibitem{771142:18601396}
L.~\&amp; Kaiser, A.~\&amp; Gomez, F.~Chollet, {Depthwise Separable
  Convolutions for Neural Machine Translation} (2017).

\bibitem{771142:18601413}
A.~G. Santos, C.~O. Souza, C.~Zanchettin, D.~Macedo, A.~L.~I. Oliveira,
  T.~Ludermir, {Reducing SqueezeNet Storage Size with Depthwise Separable
  Convolutions}, {2018 International Joint Conference on Neural Networks
  (IJCNN)} (2018) 1--6.

\bibitem{771142:18601392}
\href{https://docs.gimp.org/2.10/en/gimp-layer-white-balance.html.Accessed}{{White
  Balance}} (2020).
\newline\urlprefix\url{https://docs.gimp.org/2.10/en/gimp-layer-white-balance.html.Accessed}

\bibitem{771142:18601405}
\href{https://adadevelopment.github.io/gdal/white-balance-gdal.html}{{White
  Balance Algorithm with Gdal in C\#}} (2017).
\newline\urlprefix\url{https://adadevelopment.github.io/gdal/white-balance-gdal.html}

\bibitem{771142:18601389}
S.~M. Pizer, E.~P. Amburn, J.~D. Austin, R.~Cromartie, A.~Geselowitz, T.~Greer,
  B.~ter Haar~Romeny, J.~B. Zimmerman, K.~Zuiderveld,
  \href{https://dx.doi.org/10.1016/s0734-189x(87)80186-x}{{Adaptive histogram
  equalization and its variations}}, {Computer Vision, Graphics, and Image
  Processing} 39~(3) (1987) 355--368.
\newblock \href {https://doi.org/10.1016/s0734-189x(87)80186-x}
  {\path{doi:10.1016/s0734-189x(87)80186-x}}.
\newline\urlprefix\url{https://dx.doi.org/10.1016/s0734-189x(87)80186-x}

\bibitem{771142:18601383}
S.~M. Pizer, R.~E. Johnston, J.~P. Ericksen, B.~C. Yankaskas, K.~E. Muller,
  {Contrast-limited adaptive histogram equalization: speed and effectiveness},
  {Proceedings of the First Conference on Visualization in Biomedical
  Computing} (1990).

\bibitem{771142:18601381}
S.~Wong, Y.~Yu, N.~A. Ho, R.~Paramesran, {Comparative analysis of underwater
  image enhancement methods in different color spaces}, in: {2014 International
  Symposium on Intelligent Signal Processing and Communication Systems
  (ISPACS)}, 2014, pp. 34--038.

\bibitem{771142:18601420}
C.~Szegedy, V.~Vanhoucke, S.~Ioffe, J.~Shlens, Z.~Wojna, {Rethinking the
  Inception Architecture for Computer Vision}, {ArXiv} (2015).

\bibitem{771142:18601393}
J.~Deng, W.~Dong, R.~Socher, L.~Li, K.~Li, L.~Fei-Fei, {ImageNet: A large-scale
  hierarchical image database}, in: {2009 IEEE Conference on Computer Vision
  and Pattern Recognition}, IEEE, 2009, pp. 248--255.

\bibitem{771142:18601380}
S.~Ioffe, C.~Szegedy, {Batch Normalization: Accelerating Deep Network Training
  by Reducing Internal Covariate Shift}, {ArXiv} (2015).

\bibitem{771142:18601410}
N.~Srivastava, G.~E. Hinton, A.~Krizhevsky, I.~Sutskever, {Dropout: A simple
  way to prevent neural networks from overfitting}, {The Journal of Machine
  Learning Research} 15 (2014) 1929--1958.

\bibitem{771142:18601422}
D.~Kermany, Zhang, Kang, {Labeled Optical Coherence Tomography (OCT) and Chest
  XRay Images for Classification}, {Mendeley Data} 2 (2018).

\bibitem{771142:18601401}
C.~J. P, P.~Morrison, D.~L (2020).
\newblock \href{https://github.com/ieee8023/covid-chestxray-dataset}{[link]}.
\newline\urlprefix\url{https://github.com/ieee8023/covid-chestxray-dataset}

\bibitem{771142:18601404}
Z.-H. Chen, {Mask-RCNN detection of COVID-19 pneumonia symptoms by employing
  Stacked Autoencoders in deep unsupervised learning on Low-Dose High
  Resolution CT}, {IEEE Dataport} (2020).

\bibitem{771142:18601423}
D.~Sheet, A.~Chakravarty, T.~Sarkar, R.~Sathish, A.~Raj, V.~Balasubramanian,
  R.~Rajan, R.~Sathish, N.~Chakravorty, M.~Sinha, M.~Sharma, V.~Kumar,
  R.~Kumar, A.~Kumar, A.~Singhal, G.~Reddy, {Covid19action-radiology-CXR},
  {IEEE Dataport} (2020).

\bibitem{771142:18601424}
K.~Tang, R.~Wang, T.~Chen, {Towards Maximizing the Area Under the ROC Curve for
  Multi-Class Classification Problems}, {AAAI} (2011).

\bibitem{771142:18741313}
S.~M.~T. Ribeiro, C.~Singh, Guestrin, \href{arXiv:1602.04938}{{Marco Tulio
  Ribeiro, Sameer Singh, Carlos Guestrin."Why Should I~Trust You?": Explaining
  the Predictions of Any Classifier.arXiv:1602.04938. 2016 }} (2016).
\newline\urlprefix\url{arXiv:1602.04938}

\bibitem{771142:18601386}
R.~R. Selvaraju, M.~Cogswell, A.~Das, R.~Vedantam, D.~Parikh, D.~Batra,
  {Grad-cam: visual explanations from deep networks via gradient-based
  localization}, {Proceedings of the IEEE International Conference on Computer
  Vision. 2017}  618--626.

\bibitem{771142:18601399}
I.~D. Apostolopoulos, T.~A. Mpesiana,
  \href{https://dx.doi.org/10.1007/s13246-020-00865-4}{{Covid-19: automatic
  detection from X-ray images utilizing transfer learning with convolutional
  neural networks}}, {Physical and Engineering Sciences in Medicine} (2020).
\newblock \href {https://doi.org/10.1007/s13246-020-00865-4}
  {\path{doi:10.1007/s13246-020-00865-4}}.
\newline\urlprefix\url{https://dx.doi.org/10.1007/s13246-020-00865-4}

\bibitem{771142:18601412}
L.~Wang, A.~Wong, \href{arXiv:2003.09871}{{COVID-Net: A Tailored Deep
  Convolutional Neural Network Design for Detection of COVID-19 Cases from
  Chest Radiography Images}} (2020).
\newline\urlprefix\url{arXiv:2003.09871}

\bibitem{771142:18601411}
X.~Xu, X.~Jiang, C.~Ma, P.~Du, X.~Li, S.~Lv, \href{arXiv:2002.09334}{{Deep
  Learning System to Screen Coronavirus Disease 2019 Pneumonia. }} (2020).
\newline\urlprefix\url{arXiv:2002.09334}

\bibitem{771142:18601414}
T.~Ozturk, M.~Talo, E.~A. Yildirim, U.~B. Baloglu, O.~Yildirim, U.~R. Acharya,
  \href{https://dx.doi.org/10.1016/j.compbiomed.2020.103792}{{Automated
  detection of COVID-19 cases using deep neural networks with X-ray images}},
  {Computers in Biology and Medicine} 121 (2020) 103792--103792.
\newblock \href {https://doi.org/10.1016/j.compbiomed.2020.103792}
  {\path{doi:10.1016/j.compbiomed.2020.103792}}.
\newline\urlprefix\url{https://dx.doi.org/10.1016/j.compbiomed.2020.103792}

\bibitem{771142:18601402}
J.~Redmon, A.~Farhadi, \href{arXiv:1612.08242}{{Yolo9000: Better, Faster,
  Stronger}} (2017).
\newline\urlprefix\url{arXiv:1612.08242}

\bibitem{771142:18601421}
A.~Abbas, M.~Abdelsamea, M.~Gaber, \href{medRxiv
  2020.03.30.20047456}{{Classification of COVID-19 in chest X-ray images using
  DeTraC deep convolutional neural network}} (2020).
\newblock \href {https://doi.org/https://doi.org/10.1101/2020.03.30.20047456}
  {\path{doi:https://doi.org/10.1101/2020.03.30.20047456}}.
\newline\urlprefix\url{medRxiv 2020.03.30.20047456}

\bibitem{771142:18601406}
E.~J. Luz, P.~L. Silva, R.~Silva, L.~P. Silva, G.~J. Moreira, D.~Menotti,
  \href{arXiv:2004.05717}{{Towards an Effective and Efficient Deep Learning
  Model for COVID-19 Patterns Detection in X-ray Images}} (2020).
\newline\urlprefix\url{arXiv:2004.05717}

\bibitem{771142:18601391}
A.~Narin, C.~Kaya, Z.~Pamuk, \href{arXiv:2003.10849}{{Automatic Detection of
  Coronavirus Disease (COVID-19) Using X-Ray Images and Deep Convolutional
  Neural Networks}} (2020).
\newline\urlprefix\url{arXiv:2003.10849}

\bibitem{771142:18601382}
Y.~Song, S.~Zheng, L.~Li, X.~Zhang, X.~Zhang, Z.~Huang, Y.~{\ldots}chong,
  \href{medRxiv 2020.02.23.20026930}{{Deep learning enables accurate diagnosis
  of novel coronavirus (COVID-19) with CT images}} (2020).
\newblock \href {https://doi.org/https://doi.org/10.1101/2020.02.23.20026930}
  {\path{doi:https://doi.org/10.1101/2020.02.23.20026930}}.
\newline\urlprefix\url{medRxiv 2020.02.23.20026930}

\bibitem{771142:18601385}
S.~Wang, B.~Kang, J.~Ma, X.~Zeng, M.~Xiao, J.~Guo, \href{medRxiv
  2020.02.14.20023028}{{A deep learning algorithm using CT images to screen for
  Corona Virus Disease (COVID-19)}} (2020).
\newblock \href {https://doi.org/https://doi.org/10.1101/2020.02.14.20023028}
  {\path{doi:https://doi.org/10.1101/2020.02.14.20023028}}.
\newline\urlprefix\url{medRxiv 2020.02.14.20023028}

\bibitem{771142:18601419}
C.~Zheng, X.~Deng, Q.~Fu, Q.~Zhou, J.~Feng, H.~Ma, X.~Wang, \href{medRxiv
  2020.03.12.20027185}{{Deep learning-based detection for COVID-19 from chest
  CT using weak label}} (2020).
\newblock \href {https://doi.org/https://doi.org/10.1101/2020.03.12.20027185}
  {\path{doi:https://doi.org/10.1101/2020.03.12.20027185}}.
\newline\urlprefix\url{medRxiv 2020.03.12.20027185}

\bibitem{771142:18742186}
M.~Rahimzadeh, A.~Attar, {A modified deep convolutional neural network for
  detecting COVID-19 and pneumonia from chest X-ray images based on the
  concatenation of Xception and ResNet50V2}, {Informatics in Medicine Unlocked}
  19 (2020).
\newblock \href {https://doi.org/https://doi.org/10.1016/j.imu.2020.100360}
  {\path{doi:https://doi.org/10.1016/j.imu.2020.100360}}.

\bibitem{771142:18742357}
A.~I. Khan, J.~L. Shah, M.~Bhat, \href{arXiv:2004.04931}{{CoroNet: A Deep
  Neural Network for Detection and Diagnosis of Covid-19 from Chest X-ray
  Images}}, {arXiv} (2020).
\newline\urlprefix\url{arXiv:2004.04931}

\end{thebibliography}


\begin{thebibliography}{10}
\expandafter\ifx\csname url\endcsname\relax
  \def\url#1{\texttt{#1}}\fi
\expandafter\ifx\csname urlprefix\endcsname\relax\def\urlprefix{URL }\fi
\expandafter\ifx\csname href\endcsname\relax
  \def\href#1#2{#2} \def\path#1{#1}\fi

\bibitem{adhikari2020epidemiology}
S.~P. Adhikari, S.~Meng, Y.-J. Wu, Y.-P. Mao, R.-X. Ye, Q.-Z. Wang, C.~Sun,
  S.~Sylvia, S.~Rozelle, H.~Raat, et~al., Epidemiology, causes, clinical
  manifestation and diagnosis, prevention and control of coronavirus disease
  (covid-19) during the early outbreak period: a scoping review, Infectious
  diseases of poverty 9~(1) (2020) 1--12.

\bibitem{yi2020covid}
Y.~Yi, P.~N. Lagniton, S.~Ye, E.~Li, R.-H. Xu, Covid-19: what has been learned
  and to be learned about the novel coronavirus disease, International journal
  of biological sciences 16~(10) (2020) 1753.

\bibitem{udugama2020diagnosing}
B.~Udugama, P.~Kadhiresan, H.~N. Kozlowski, A.~Malekjahani, M.~Osborne, V.~Y.
  Li, H.~Chen, S.~Mubareka, J.~B. Gubbay, W.~C. Chan, Diagnosing covid-19: the
  disease and tools for detection, ACS nano 14~(4) (2020) 3822--3835.

\bibitem{kanne2020essentials}
J.~P. Kanne, B.~P. Little, J.~H. Chung, B.~M. Elicker, L.~H. Ketai, Essentials
  for radiologists on covid-19: an update—radiology scientific expert panel
  (2020).

\bibitem{xie2020chest}
X.~Xie, Z.~Zhong, W.~Zhao, C.~Zheng, F.~Wang, J.~Liu, Chest ct for typical
  2019-ncov pneumonia: relationship to negative rt-pcr testing, Radiology
  (2020) 200343.

\bibitem{kong2020chest}
W.~Kong, P.~P. Agarwal, Chest imaging appearance of covid-19 infection,
  Radiology: Cardiothoracic Imaging 2~(1) (2020) e200028.

\bibitem{lee2020covid}
E.~Y. Lee, M.-Y. Ng, P.-L. Khong, Covid-19 pneumonia: what has ct taught us?,
  The Lancet Infectious Diseases 20~(4) (2020) 384--385.

\bibitem{shi2020radiological}
H.~Shi, X.~Han, N.~Jiang, Y.~Cao, O.~Alwalid, J.~Gu, Y.~Fan, C.~Zheng,
  Radiological findings from 81 patients with covid-19 pneumonia in wuhan,
  china: a descriptive study, The Lancet Infectious Diseases (2020).

\bibitem{zhao2020relation}
W.~Zhao, Z.~Zhong, X.~Xie, Q.~Yu, J.~Liu, Relation between chest ct findings
  and clinical conditions of coronavirus disease (covid-19) pneumonia: A
  multicenter study, American Journal of Roentgenology 214 (2020) 1--6.
\newblock \href {https://doi.org/10.2214/AJR.20.22976}
  {\path{doi:10.2214/AJR.20.22976}}.

\bibitem{li2020coronavirus}
Y.~Li, L.~Xia, Coronavirus disease 2019 (covid-19): role of chest ct in
  diagnosis and management, American Journal of Roentgenology (2020) 1--7.

\bibitem{yoon2020chest}
S.~H. Yoon, K.~H. Lee, J.~Y. Kim, Y.~K. Lee, H.~Ko, K.~H. Kim, C.~M. Park,
  Y.-H. Kim, Chest radiographic and ct findings of the 2019 novel coronavirus
  disease (covid-19): analysis of nine patients treated in korea, Korean
  journal of radiology 21~(4) (2020) 494--500.

\bibitem{stephen2019efficient}
O.~Stephen, M.~Sain, U.~J. Maduh, D.-U. Jeong, An efficient deep learning
  approach to pneumonia classification in healthcare, Journal of healthcare
  engineering 2019 (2019).

\bibitem{chhikara2020deep}
P.~Chhikara, P.~Singh, P.~Gupta, T.~Bhatia, Deep convolutional neural network
  with transfer learning for detecting pneumonia on chest x-rays, in: Advances
  in Bioinformatics, Multimedia, and Electronics Circuits and Signals,
  Springer, 2020, pp. 155--168.

\bibitem{sethy2020detection}
P.~K. Sethy, S.~K. Behera, Detection of coronavirus disease (covid-19) based on
  deep features, Preprints 2020030300 (2020) 2020.

\bibitem{loey2020within}
M.~Loey, F.~Smarandache, N.~E. M~Khalifa, Within the lack of chest covid-19
  x-ray dataset: A novel detection model based on gan and deep transfer
  learning, Symmetry 12~(4) (2020) 651.

\bibitem{hemdan2020covidx}
E.~E.-D. Hemdan, M.~A. Shouman, M.~E. Karar, Covidx-net: A framework of deep
  learning classifiers to diagnose covid-19 in x-ray images, arXiv preprint
  arXiv:2003.11055 (2020).

\bibitem{hazra2018robust}
S.~Hazra, A.~Santra, Robust gesture recognition using millimetric-wave radar
  system, IEEE sensors letters 2~(4) (2018) 1--4.

\bibitem{picone1993signal}
J.~W. Picone, Signal modeling techniques in speech recognition, Proceedings of
  the IEEE 81~(9) (1993) 1215--1247.

\bibitem{chollet2017xception}
F.~Chollet, Xception: Deep learning with depthwise separable convolutions, in:
  Proceedings of the IEEE conference on computer vision and pattern
  recognition, 2017, pp. 1251--1258.

\bibitem{whitebal2020}
\href{https://docs.gimp.org/2.10/en/gimp-layer-white-balance.html.Accessed}{{White
  Balance}} (2020).
\newline\urlprefix\url{https://docs.gimp.org/2.10/en/gimp-layer-white-balance.html.Accessed}

\bibitem{whitebal2020gdal}
\href{https://adadevelopment.github.io/gdal/white-balance-gdal.html}{{White
  Balance Algorithm with Gdal in C\#}} (2017).
\newline\urlprefix\url{https://adadevelopment.github.io/gdal/white-balance-gdal.html}

\bibitem{pizer1987adaptive}
S.~M. Pizer, E.~P. Amburn, J.~D. Austin, R.~Cromartie, A.~Geselowitz, T.~Greer,
  B.~ter Haar~Romeny, J.~B. Zimmerman, K.~Zuiderveld, Adaptive histogram
  equalization and its variations, Computer vision, graphics, and image
  processing 39~(3) (1987) 355--368.

\bibitem{pizer1990contrast}
S.~M. Pizer, Contrast-limited adaptive histogram equalization: Speed and
  effectiveness stephen m. pizer, r. eugene johnston, james p. ericksen, bonnie
  c. yankaskas, keith e. muller medical image display research group, in:
  Proceedings of the First Conference on Visualization in Biomedical Computing,
  Atlanta, Georgia, May 22-25, 1990, IEEE Computer Society Press, 1990, p. 337.

\bibitem{wong2014comparative}
S.-L. Wong, Y.-P. Yu, N.~A.-J. Ho, R.~Paramesran, Comparative analysis of
  underwater image enhancement methods in different color spaces, in: 2014
  International Symposium on Intelligent Signal Processing and Communication
  Systems (ISPACS), IEEE, 2014, pp. 034--038.

\bibitem{szegedy2016rethinking}
C.~Szegedy, V.~Vanhoucke, S.~Ioffe, J.~Shlens, Z.~Wojna, Rethinking the
  inception architecture for computer vision, in: Proceedings of the IEEE
  conference on computer vision and pattern recognition, 2016, pp. 2818--2826.

\bibitem{deng2009imagenet}
J.~Deng, W.~Dong, R.~Socher, L.-J. Li, K.~Li, L.~Fei-Fei, Imagenet: A
  large-scale hierarchical image database, in: 2009 IEEE conference on computer
  vision and pattern recognition, Ieee, 2009, pp. 248--255.

\bibitem{ioffe2015batch}
S.~Ioffe, C.~Szegedy, Batch normalization: Accelerating deep network training
  by reducing internal covariate shift, arXiv preprint arXiv:1502.03167 (2015).

\bibitem{srivastava2014dropout}
N.~Srivastava, G.~Hinton, A.~Krizhevsky, I.~Sutskever, R.~Salakhutdinov,
  Dropout: a simple way to prevent neural networks from overfitting, The
  journal of machine learning research 15~(1) (2014) 1929--1958.

\bibitem{kermany2018large}
D.~Kermany, K.~Zhang, M.~Goldbaum, Large dataset of labeled optical coherence
  tomography (oct) and chest x-ray images, Mendeley Data, v3 http://dx. doi.
  org/10.17632/rscbjbr9sj 3 (2018).

\bibitem{cohen2020data}
C.~J. P, P.~Morrison, D.~L,
  \href{https://github.com/ieee8023/covid-chestxray-dataset}{Covid-19 image
  data collection, arxiv}, arXiv preprint arXiv:2003.11597 (2020).
\newline\urlprefix\url{https://github.com/ieee8023/covid-chestxray-dataset}

\bibitem{4kcm-m312-20}
Z.-H. Chen, \href{http://dx.doi.org/10.21227/4kcm-m312}{Mask-rcnn detection of
  covid-19 pneumonia symptoms by employing stacked autoencoders in deep
  unsupervised learning on low-dose high resolution ct} (2020).
\newblock \href {https://doi.org/10.21227/4kcm-m312}
  {\path{doi:10.21227/4kcm-m312}}.
\newline\urlprefix\url{http://dx.doi.org/10.21227/4kcm-m312}

\bibitem{dsheetieee}
D.~S. et.al,
  \href{http://dx.doi.org/10.21227/s7pw-jr18}{Covid19action-radiology-cxr}
  (2020).
\newblock \href {https://doi.org/10.21227/s7pw-jr18}
  {\path{doi:10.21227/s7pw-jr18}}.
\newline\urlprefix\url{http://dx.doi.org/10.21227/s7pw-jr18}

\bibitem{tang2011towards}
K.~Tang, R.~Wang, T.~Chen, Towards maximizing the area under the roc curve for
  multi-class classification problems, in: Twenty-Fifth AAAI Conference on
  Artificial Intelligence, 2011.

\bibitem{ribeiro2016should}
M.~T. Ribeiro, S.~Singh, C.~Guestrin, " why should i trust you?" explaining the
  predictions of any classifier, in: Proceedings of the 22nd ACM SIGKDD
  international conference on knowledge discovery and data mining, 2016, pp.
  1135--1144.

\bibitem{selvaraju2017grad}
R.~R. Selvaraju, M.~Cogswell, A.~Das, R.~Vedantam, D.~Parikh, D.~Batra,
  Grad-cam: Visual explanations from deep networks via gradient-based
  localization, in: Proceedings of the IEEE international conference on
  computer vision, 2017, pp. 618--626.

\bibitem{jacobi2020portable}
A.~Jacobi, M.~Chung, A.~Bernheim, C.~Eber, Portable chest x-ray in coronavirus
  disease-19 (covid-19): A pictorial review, Clinical Imaging 64 (04 2020).
\newblock \href {https://doi.org/10.1016/j.clinimag.2020.04.001}
  {\path{doi:10.1016/j.clinimag.2020.04.001}}.

\bibitem{wong2020frequency}
H.~Y.~F. Wong, H.~Y.~S. Lam, A.~H.-T. Fong, S.~T. Leung, T.~W.-Y. Chin,
  C.~S.~Y. Lo, M.~M.-S. Lui, J.~C.~Y. Lee, K.~W.-H. Chiu, T.~Chung, et~al.,
  Frequency and distribution of chest radiographic findings in covid-19
  positive patients, Radiology (2020) 201160.

\bibitem{apostolopoulos2020covid}
I.~D. Apostolopoulos, T.~A. Mpesiana, Covid-19: automatic detection from x-ray
  images utilizing transfer learning with convolutional neural networks,
  Physical and Engineering Sciences in Medicine (2020) 1.

\bibitem{wang2020covid}
L.~Wang, A.~Wong, Covid-net: A tailored deep convolutional neural network
  design for detection of covid-19 cases from chest radiography images, arXiv
  (2020) arXiv--2003.

\bibitem{xu2020others}
X.~Xu, X.~Jiang, C.~Ma, P.~Du, X.~Li, S.~Lv, L.~Yu, Y.~Chen, J.~Su, G.~Lang,
  others," deep learning system to screen coronavirus disease 2019 pneumonia,",
  arXiv preprint arXiv:2002.09334 (2020).

\bibitem{ozturk2020automated}
T.~Ozturk, M.~Talo, E.~A. Yildirim, U.~B. Baloglu, O.~Yildirim, U.~R. Acharya,
  Automated detection of covid-19 cases using deep neural networks with x-ray
  images, Computers in Biology and Medicine (2020) 103792.

\bibitem{redmon2017yolo9000}
J.~Redmon, A.~Farhadi, Yolo9000: Better, faster, stronger, arXiv (2016).

\bibitem{abbas2020classification}
A.~Abbas, M.~Abdelsamea, M.~Gaber, Classification of covid-19 in chest x-ray
  images using detrac deep convolutional neural network, medRxiv.

\bibitem{luz2020towards}
E.~Luz, P.~L. Silva, R.~Silva, G.~Moreira, Towards an efficient deep learning
  model for covid-19 patterns detection in x-ray images, arXiv preprint
  arXiv:2004.05717 (2020).

\bibitem{narin2020automatic}
A.~Narin, C.~Kaya, Z.~Pamuk, Automatic detection of coronavirus disease
  (covid-19) using x-ray images and deep convolutional neural networks, arXiv
  preprint arXiv:2003.10849 (2020).

\bibitem{song2020deep}
Y.~Song, S.~Zheng, L.~Li, X.~Zhang, X.~Zhang, Z.~Huang, J.~Chen, H.~Zhao,
  Y.~Jie, R.~Wang, et~al., Deep learning enables accurate diagnosis of novel
  coronavirus (covid-19) with ct images, medRxiv (2020).

\bibitem{wang2020deep}
S.~Wang, B.~Kang, J.~Ma, X.~Zeng, M.~Xiao, J.~Guo, M.~Cai, J.~Yang, Y.~Li,
  X.~Meng, et~al., A deep learning algorithm using ct images to screen for
  corona virus disease (covid-19), MedRxiv (2020).

\bibitem{zheng2020deep}
C.~Zheng, X.~Deng, Q.~Fu, Q.~Zhou, J.~Feng, H.~Ma, W.~Liu, X.~Wang, Deep
  learning-based detection for covid-19 from chest ct using weak label, medRxiv
  (2020).

\bibitem{rahimzadeh2020modified}
M.~Rahimzadeh, A.~Attar, A modified deep convolutional neural network for
  detecting covid-19 and pneumonia from chest x-ray images based on the
  concatenation of xception and resnet50v2, Informatics in Medicine Unlocked
  (2020) 100360\href {https://doi.org/10.1016/j.imu.2020.100360}
  {\path{doi:10.1016/j.imu.2020.100360}}.

\bibitem{khan2020coronet}
A.~I. Khan, J.~L. Shah, M.~Bhat, Coronet: A deep neural network for detection
  and diagnosis of covid-19 from chest x-ray images, arXiv preprint
  arXiv:2004.04931 (2020).

\end{thebibliography}

\end{document}